\documentclass[12pt]{article}
\usepackage{amssymb,amsmath,mathrsfs}
\usepackage{hyperref}
\usepackage{graphicx}
\usepackage{color}
\usepackage[OT2,OT1]{fontenc}
\usepackage{upquote}
\usepackage{authblk}
\usepackage[numbers,sort&compress]{natbib}

\begin{document}

\title{\LARGE\bf Rapid computation of the total band radiance by using the Spectrally Integrated Voigt Function}

\bigskip
\author[1, 2, 3]{\small Sanjar M. Abrarov}
\author[1, 2, 3, 4]{\small Rehan Siddiqui}
\author[2, 3, 4]{\small Rajinder K. Jagpal}
\author[1, 4]{\small \\ Brendan M. Quine}

\affil[1]{\scriptsize Dept. Earth and Space Science and Engineering, York University, 4700 Keele St., Canada, M3J 1P3 \normalsize}
\affil[2]{\scriptsize Epic Climate Green (ECG) Inc., 23 Westmore Dr., Unit 310, Toronto, M339V 3Y7 \normalsize}
\affil[3]{\scriptsize Epic College of Technology, 5670 McAdam Rd., Mississauga, Canada, L4Z 1T2 \normalsize}
\affil[4]{\scriptsize Dept. Physics and Astronomy, York University, 4700 Keele St., Toronto, Canada, M3J 1P3 \normalsize}

\date{February 15, 2021}
\maketitle
%\vspace{1cm}
%\bigskip

\begin{abstract}
In our earlier publication we introduced the Spectrally Integrated Voigt Function (SIVF) as an alternative to the traditional Voigt function for the HITRAN-based applications \href{https://doi.org/10.1016/j.jqsrt.2013.04.020}{[Quine \& Abrarov, JQSRT 2013]}. It was shown that application of the SIVF enables us to reduce spectral resolution without loss of accuracy in computation of the spectral radiance. As a further development, in this study we present more efficient SIVF approximations derived by using a new sampling method based on incomplete cosine expansion of the sinc function \href{https://doi.org/10.1016/j.amc.2015.01.072}{[Abrarov \& Quine, Appl. Math. Comput. 2015]}. Since the SIVF mathematically represents the mean value integral of the Voigt function, this method accounts for area under the curve of the Voigt function. Consequently, the total band radiance, defined as the integrated spectral radiance within a given spectral region, can also retain its accuracy even at low spectral resolution. Our numerical results demonstrate that application of the SIVF may be promising for more rapid line-by-line computation in atmospheric models utilizing the HITRAN molecular spectroscopic database. Such an approach may be particularly efficient to implement a retrieval algorithm for the greenhouse gases from the NIR space data collected by Earth-orbiting micro-spectrometers like Argus 1000 for their operation in a real-time mode. The real-time mode operation of the micro-spectrometers can be advantageous for instant decision making during flight for more efficient data collection from space.
\vspace{0.25cm}
\vspace{0.25cm}
\end{abstract}

\noindent{\bf\large{Nomenclature}} \\
$_j$ - HITRAN subscript for gas specie \\
$_i$ - HITRAN subscript for line center \\

\noindent{\bf\large{Acronyms/Abbreviations}} \\
SIVF – Spectrally Integrated Voigt Function \\
LBL – line-by line

\section{Introduction}
In 1992 Brüggemann and Bollig introduced an integral \cite{Bruggemann1992a}
\begin{equation}\label{eq_1}
I\left( {x,y} \right) = \int {K\left( {x,y} \right)dx},
\end{equation}
where $K\left( {x,y} \right)$ is the Voigt function (also known as the Voigt distribution) representing a convolution of the Lorenz and Doppler functions. Specifically, using the Gauss–Hermite quadrature they showed that the function $I\left( {x,y} \right)$ can be represented in a logarithmic form \cite{Bruggemann1992a} (see also \cite{Quine2013} for more details)
\begin{equation}\label{eq_2}
I\left( {x,y} \right) \approx {\mathop{\rm Re}\nolimits} \left[ {\frac{i}{\pi }\ln \prod\limits_{r = 1}^{n/2} {{{\left\{ {{{\left( {x + iy} \right)}^2} + {{\left( {x_r^{\left\langle n \right\rangle }} \right)}^2}} \right\}}^{\lambda _r^{\left\langle n \right\rangle }}}} } \right] + \,\,\frac{{\sqrt \pi  }}{2},
\end{equation}
where $x_r^{\left\langle n \right\rangle }$ and $\lambda _r^{\left\langle n \right\rangle }$ are roots and weights of the Hermite polynomial of the order $n$, respectively. 

Brüggemann and Bollig in \cite{Bruggemann1992a} considered only derivation and some computational aspects of the integral \eqref{eq_1} without any suggestion of how this function can be applied in the Atmospheric Science or elsewhere in spectroscopy. However, this function has appeared to be interesting and very promising for spectroscopic applications \cite{Quine2013, Ilakovac2019}. Although Brüggemann in his subsequent publication \cite{Bruggemann1992b} stated that the Voigt distribution can be used in computation of the total integrated intensity, it remains unclear how the function $I\left( {x,y} \right)$ rather than the Voigt distribution itself can be utilized in calculation of the total integrated intensity.

The approximation \eqref{eq_2} has some limitation as its accuracy deteriorates at $y < 0.5$ (see Table 1 in \cite{Quine2013}). Furthermore, it is not rapid enough in computation. Therefore, we attempted to find alternative approximations for the function $I\left( {x,y} \right)$ by taking numerically the integral \eqref{eq_1} (see equations (11), (15) and (17) in our paper \cite{Quine2013}). Although all these three approximations are accurate, the application of the equation (11) from our earlier work \cite{Quine2013} is not fast enough due to dependence of the expansion coefficient $b\left( x \right)$ on the parameter $x$. This motivated us to find other approximations for rapid and accurate computation of the function $I\left( {x,y} \right)$.

The Voigt function that can be defined as \cite{Armstrong1967, Srivastava1987, Srivastava1992, Abrarov2011}
\begin{equation}\label{eq_3}
K\left( {x,y} \right) = \frac{1}{{\sqrt 2 }}\int\limits_0^\infty  {{e^{ - {t^2}/4}}{e^{ - yt}}\cos \left( {xt} \right)dt, \qquad y \ge 0},
\end{equation}
is widely used in many applications of the radiative transfer based on HITRAN molecular spectroscopic database \cite{Hill2016}. The input parameters of the Voigt function are given by
$$
x = \frac{{\nu - {\nu_i}}}{{{\alpha _D}}}\sqrt {\ln 2}
$$
and
$$
y = \frac{{{\alpha _L}}}{{{\alpha _D}}}\sqrt {\ln 2},
$$
where $\nu$ is the frequency $\left[cm^{-1}\right]$, ${\nu_i}$ the line center frequency $\left[cm^{-1}\right]$, ${\alpha _L}$ and ${\alpha _D}$ are Lorentz and Doppler half widths (HWHMs). All required spectroscopic data of the Voigt profile
$$
V\left( {x,y} \right) = {K_0}K\left( {x,y} \right),
$$
where
$$
{K_0} = \frac{1}{{{\alpha _D}}}\sqrt {\ln 2},
$$
are provided by the HITRAN database \cite{Hill2016}.

Since its introduction in \cite{Bruggemann1992a} the integral \eqref{eq_1} had not been applied in practice for more than three decades. In our earlier publication \cite{Quine2013} we proposed a first practical application of the function $I\left( {x,y} \right)$ for LBL radiative transfer modeling \cite{Edwards1988, Edwards1992, Scott1981, Clough1992, Quine2002, Berk2017}. Recently, Ilakovac suggested another interesting application of the function $I\left( {x,y} \right)$ in analysis of experimental line spectra \cite{Ilakovac2019}.

In our work \cite{Quine2013} we introduced the SIVF as
\begin{equation}\label{eq_4}
\bar K\left( {x,y,\Delta x} \right) = \frac{1}{{\Delta x}}\int\limits_{x - \Delta x/2}^{x + \Delta x/2} {K\left( {x',y} \right)dx'},
\end{equation}
where $\Delta x = {x_{n + 1}} - {x_n}$ is the difference between two adjacent grid-points and implemented SIVF in the LBL forward model GENSPECT \cite{Quine2002}. Comparing equations \eqref{eq_1} with \eqref{eq_4} one can see that SIVF can be expressed as
\begin{equation}\label{eq_5}
\begin{aligned}
\bar K\left( {x,y,\Delta x} \right) &= \left. {\frac{1}{{\Delta x}}I\left( {x',y} \right)} \right|_{x - \Delta x/2}^{x + \Delta x/2}\\
 &= \frac{1}{{\Delta x}}\left( {I\left( {x + \frac{{\Delta x}}{2},y} \right) - I\left( {x - \frac{{\Delta x}}{2},y} \right)} \right).
\end{aligned}
\end{equation}

As it has been shown in our publication \cite{Quine2013}, the application of the SIVF may be more advantageous as compared to the conventional Voigt function \eqref{eq_3}. The SIVF represents the mean value of the Voigt function and due to integration with respect to the parameter $x$ dependent to the frequency $v$, the mean value of the Voigt function accounts for the area under the curve bounded within the range between grid-points ${x_n}$ and ${x_{n + 1}}$. Consequently, it is mathematically more informative as compared to the conventional Voigt function \eqref{eq_3}. In fact, the SIVF can be transformed into the Voigt function by using the following limit
\[
K\left( {x,y} \right)  = \mathop{\lim}\limits_{\Delta x \to 0}\bar K(x,y,\Delta x) = \mathop {\lim }\limits_{\Delta x \to 0 } \frac{1}{{\Delta x}}\int\limits_{x - \Delta x/2}^{x + \Delta x/2} {K\left( {x',y} \right)dx'}.
\]
The computational test we performed shows that at $\Delta x \le {10^{ - 6}}$ the SIVF and Voigt function provide practically same values at given input arguments $x$ and $y$.

\begin{figure}[ht]
\begin{center}
\includegraphics[width=20pc]{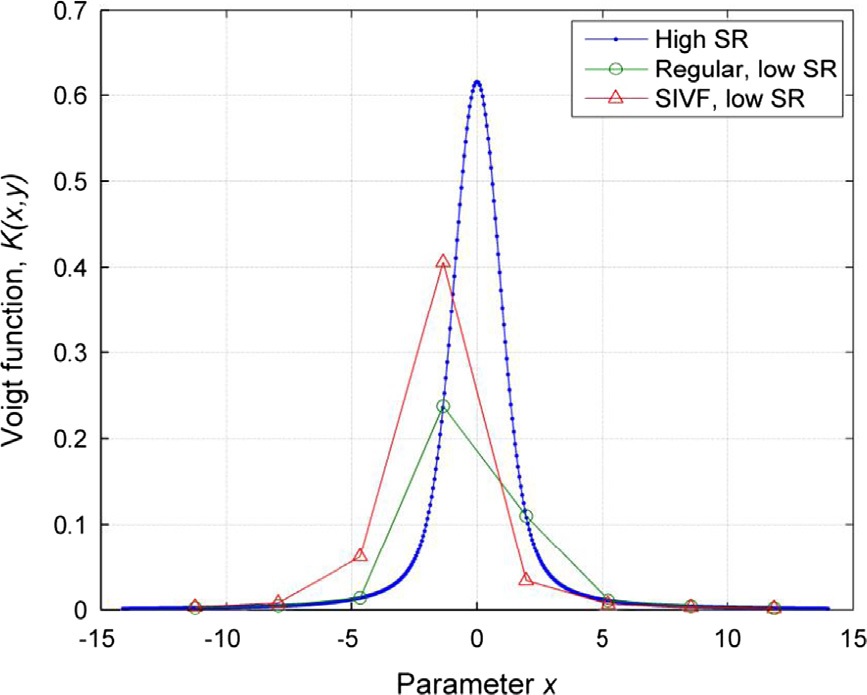}\hspace{2pc}%
\begin{minipage}[b]{22pc}
\vspace{0.3cm}
\begin{center}
{\sffamily {\bf{Fig. 1.}}  The Voigt and SIVF functions.}
\end{center}
\end{minipage}
\end{center}
\end{figure}

Figure 1 shows the Voigt function computed at high (blues curve) and low (green curve) spectral resolutions, respectively. The red curve shows the SIVF computed at low spectral resolution. As we can see, the green curve “cuts” a significant portion of the curve and, consequently, its absolute value is greatly underestimated. Although the value of the red cure is also underestimated, it always preserves the area under the curve regardless the magnitude of spectral resolution. This can be seen from the Fig. 2 showing the Voigt function (blue curve) and SIVF computed at low spectral resolution (red curve). More specifically, the area under the curve of the Voigt function within any spectral range $\Delta x = {x_{n + 1}} - {x_n}$ is absolutely equal to the area under the curve of SIVF. The black stepwise function in the Fig. 2 is a rearranged SIVF curve that is shown to emphasize its area for each interval from ${x_n}$ to ${x_{n + 1}}$. Since SIVF preserves the area under the curve, it can be used to compute the total band radiance (or intensity per steradian) at low spectral resolution more efficiently than the regular Voigt function \eqref{eq_3}. The requirement for the low spectral resolution is to accelerate computation in a LBL radiative transfer model.

\begin{figure}[ht]
\begin{center}
\includegraphics[width=20pc]{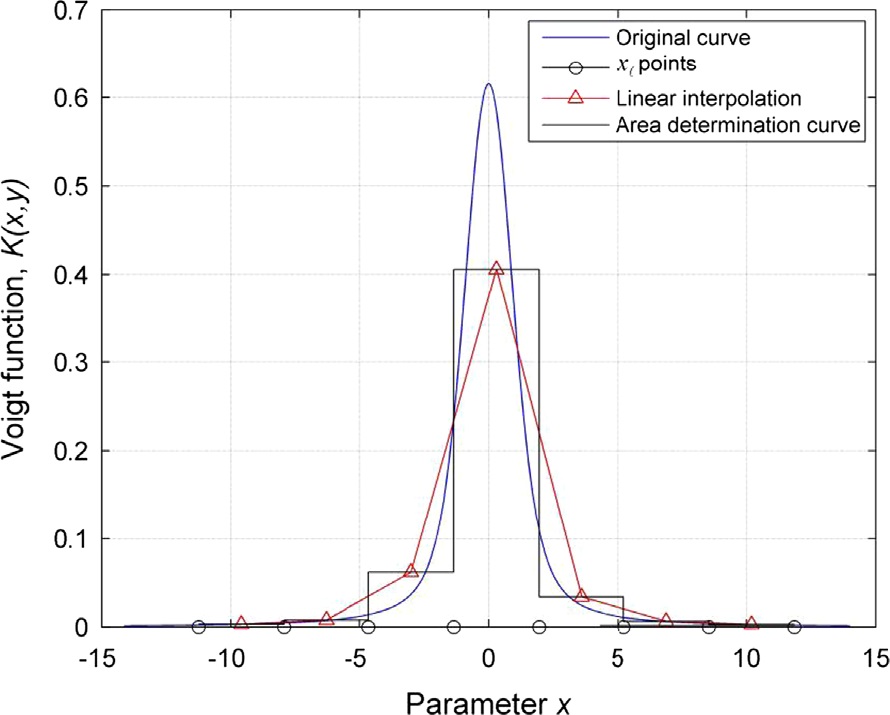}\hspace{2pc}%
\begin{minipage}[b]{22pc}
\vspace{0.3cm}
{\sffamily {\bf{Fig. 2.}} The Voigt function and the area under its curve.}
\end{minipage}
\end{center}
\end{figure}

In our earlier publication \cite{Quine2013} we have shown that application of the SIVF can be used to reduce spectral resolution without loss of accuracy. In particular, we have shown that at moderate spectral resolution with step in grid-points ${\nu_{n + 1}} - {\nu_n} = \Delta \nu = 0.005\,c{m^{ - 1}}$ the SIVF provides practically same accuracy in computation of the spectral radiance as the regular Voigt function at high spectral resolution with step in grid-points $\Delta \nu = 0.00025\,c{m^{ - 1}}$ \cite{Quine2013}. As a further development, in this work we derive the new approximations for the SIVF and show how this approach can also be implemented at a low spectral resolution $\Delta \nu = 0.1\,c{m^{ - 1}}$ to accelerate the computation of the total band radiance.

\section{Methodology}

The spectral radiance [$W{m^{ - 2}}s{r^{ - 1}}{\left( {c{m^{ - 1}}} \right)^{ - 1}}$], can be computed by using the following equation \cite{Edwards1992, Liou2002}
\begin{equation}\label{eq_6}
\begin{aligned}
{R_{sp}}\left( {\nu ,{z_{obs}}} \right) =& {R_{sp}}\left( {\nu ,{z_s}} \right){\cal T}\left( {\nu ,{z_s},{z_{obs}}} \right)\\
 &  + \int\limits_{{z_s}}^{{z_{obs}}} {B\left( {\nu ,T\left( z \right)} \right){\cal T}\left( {\nu ,{z_s},{z_{obs}}} \right)k\left( {\nu ,z} \right)\rho \left( z \right)dz} ,
\end{aligned}
\end{equation}
where ${R_{sp}}\left( {\nu ,{z_s}} \right)$ is the spectral radiance from the source point ${z_s}$, ${\cal T}\left( {\nu ,{z_s},{z_{obs}}} \right)$ is the optical transmittance from the source ${z_s}$ to observation point ${z_{obs}}$ and $B\left( {\nu ,T\left( z \right)} \right)$ is the blackbody radiation depending of frequency $\nu$ and temperature $T$.

The optical transmittance in equation \eqref{eq_6} is dependent on the absorption coefficient since
$${\cal T}\left( {\nu ,{z_s},{z_{obs}}} \right) =  - \int\limits_{{z_s}}^{{z_{obs}}} {k\left( {\nu ,z} \right)\rho \left( z \right)dz},
$$
where $\rho \left( z \right)$ is the atmospheric gas density and
$$
k\left( \nu  \right) = \sum\limits_j {{k_j}\left( \nu \right)},
$$
is the total absorption coefficient due to contribution of each atmospheric gas specie $j$. Computation of the spectral radiance involves the Voigt profile since the absorption coefficient of a gas $j$ is given by \cite{Hill2016, Quine2002}
$$
{k_j}\left( \nu  \right) = \sum\limits_i {{S_i}_jV\left( {x\left( {\nu,{\nu_i},{\alpha _D}} \right),y\left( {{\alpha _L},{\alpha _D}} \right)} \right)},
$$
where ${S_i}_j$ is the strength. All required parameters used in computation ${\nu_i}$, ${\alpha _L}$, ${\alpha _D}$ and ${S_i}_j$ are provided by the HITRAN molecular spectroscopic database \cite{Hill2016}.

In the LBL radiative transfer model like GENSPECT \cite{Quine2002}, the computations can be performed by slicing a gas column to number of the cells such that each gas cell can be considered as a homogeneous gas medium. Although this approach can efficiently overcomes many problems related to inhomogeneity of a gas medium, it requires intense computation that may take a considerable amount of time especially at higher spectral resolution since each gas has to be computed separately for each individual cell.

Rapid computation is especially desirable for space-orbiting remote sensors like Argus 1000 \cite{Jagpal2010, Jagpal2019} and Argus 2000 \cite{Issa2020, Jallad2019}. Particularly, a space-orbiting micro-spectrometer may require operation in a real-time mode for instant decision making to change its parameters such as snap-shot imaging frequency, integration time, deviation of nadir angle, etc. \cite{Quine2014} in order to take a probe over a specific area on the ground for a source of the pollutants like increased concentration of $\rm{CO_2}$ gas or combustion aerosols that may appear due to industrial activities, seasonal forest or field fires. Furthermore, due to high velocity of a space-orbiting micro-spectrometer that is above the first cosmic velocity $7.8\,km/s$, the decision making time should not exceed a few milliseconds to insure that the probing area on the ground has not been missed during the flight.

One of the most efficient ways to resolve this problem is to decrease the spectral resolution. However, the decreased spectral resolution causes some side effects including distortion of the shape of the spectral band and missed individual lines that occur as a result of the “cuts” of the spikes of the Voigt function (see Fig. 1). Therefore, the lower bound of resolution in computation utilizing the conventional Voigt function is very limited. However, in contrast to the conventional Voigt function the SIVF significantly better preserves the shape of the spectral band and does not miss individual lines. In particular, at moderate spectral resolution $0.005\,c{m^{ - 1}}$ the application of SIVF perfectly retains accuracy in the spectral radiance (see Figs. 12, 13 and 14 in \cite{Quine2013}). In this work we show that the SIVF can also be efficient in computation of the total band radiance (intensity per steradian [$W{m^{ - 2}}s{r^{ - 1}}$]) defined as
\begin{equation}\label{eq_7}
TBR = \int\limits_{{\nu_{\min }}}^{{\nu _{\max }}} {{R_{sp}}\left( {\nu ,{z_{obs}}} \right)d\nu },
\end{equation}
where ${\nu_{\min }}$ and ${\nu_{\max }}$ are the lower and upper bounds of integration, such that ${\nu_{\max }} - {\nu_{\min }} = \Delta {\nu_{{\rm{band}}}}$ is the integration band over the spectral radiance ${R_{sp}}\left( {\nu ,{z_{obs}}} \right)$. In particular, the computational tests we performed for $\rm{CO_2}$ greenhouse gas show that the total band radiance can retain accuracy even at a low spectral resolution \cite{Quine2013}.

It should be noted that communication of the space-orbiting micro- spectrometer with ground stations causes a significant delay up to one – two seconds. Therefore, the SIVF methodology may be particularly efficient for on-board computing in a spacecraft for instant data retrieval, analysis and decision making due to a significantly lower spectral resolution requirement.

\section{Derivation}

\subsection{Domain scheme}

The domain scheme for rapid and accurate computation of the SIVF is shown in Fig. 3. It consists of external and internal domains, where internal domains are subdivided into main and supplementary internal subdomains. All applications based on the HITRAN require the parameter $y > 0.$ Furthermore, since $I\left( {x,y} \right)$ is an even (symmetric) function, we can consider only the first quadrant of the plane. The main internal domain is bounded by set of numbers  $\left| {x + iy} \right| < 15 \cap \left| x \right| \ge {10^{ - 6}}$. Consequently, the supplementary internal domain is bounded by set of numbers$\left| {x + iy} \right| < 15 \cap \left| x \right| < {10^{ - 6}}$. The external domain is bounded by set of numbers $\left| {x + iy} \right| \ge 15$.

\begin{figure}[ht]
\begin{center}
\includegraphics[width=20pc]{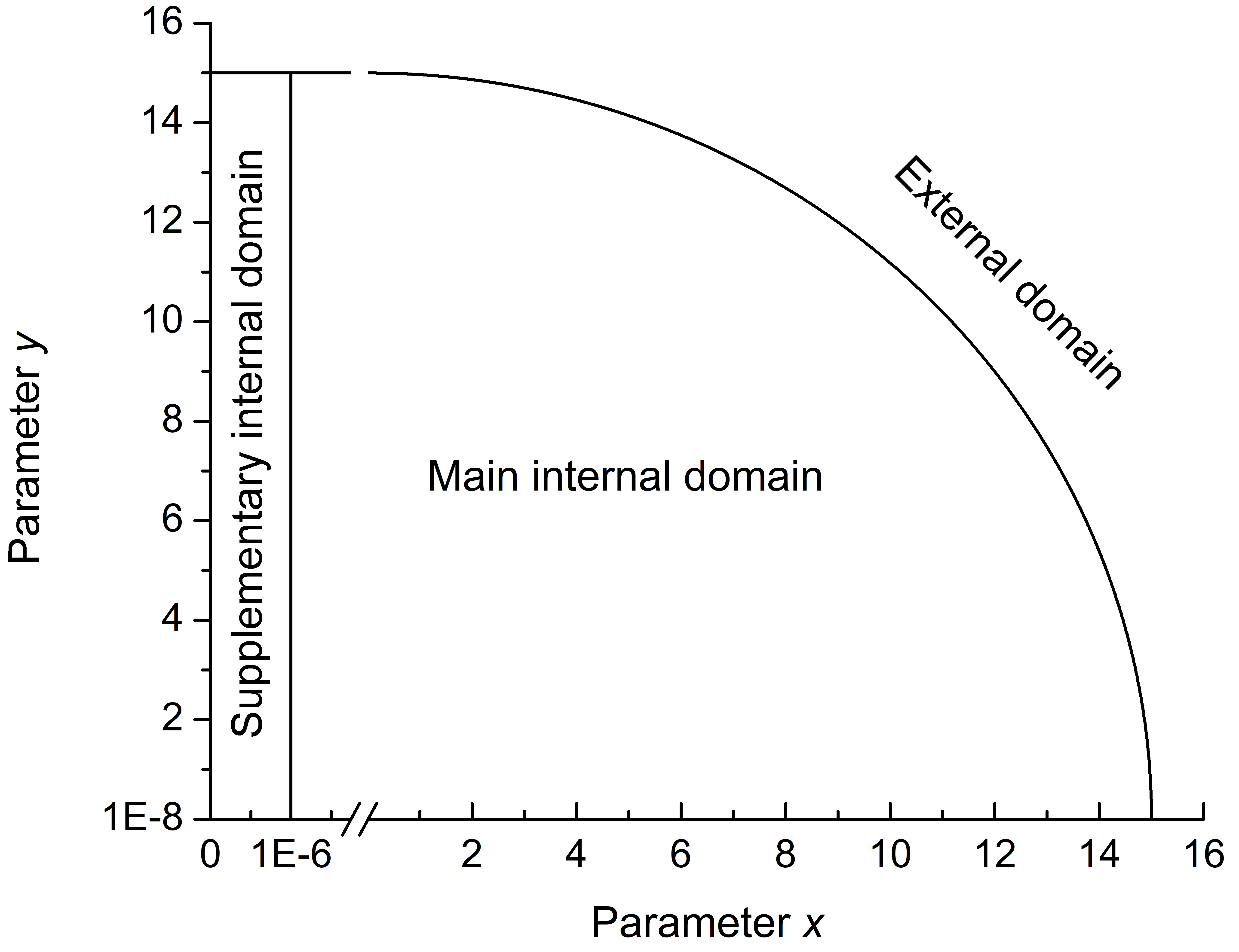}\hspace{2pc}%
\begin{minipage}[b]{22pc}
\vspace{0.3cm}
{\sffamily {\bf{Fig. 3.}} 3-domain scheme for SIVF implementation.}
\end{minipage}
\end{center}
\end{figure}

\subsection{External domain}

Substituting equation \eqref{eq_3} into \eqref{eq_1} leads to \cite{Quine2013}
$$
\int {K\left( {x,y} \right)dx}  = \frac{x}{{\sqrt \pi  }}\int\limits_0^\infty  {{{\mathop{\rm e}\nolimits} ^{ - {t^2}/4}}{{\mathop{\rm e}\nolimits} ^{ - yt}}{\rm{sinc}}\left( {xt} \right)dt}
$$
and we can see that at larger values $x$ or $y$ the integrand on the right side integral rapidly tends to zero with increasing $t$. Consequently, at $\left| {x + iy} \right| \ge 15$ only small initial part of the integrand contributes for integration. As a result, for the external domain it is sufficient to approximate the exponential function ${e^{ - {t^2}/4}}$ just over a very narrow interval $t \in \left[ {0,{t_0}} \right]$, such that ${t_0} << 1$. Thus, by using the relation that we found by empirical match
$$
{e^{ - {t^2}/4}} \approx \cos \left( {t/\sqrt 2 } \right),	\qquad t \in \left[ {0,{t_0}} \right], \qquad t_0 << 1,
$$
and taking into account that
$$
\int {\int\limits_0^\infty  {{e^{ - {t^2}/4}}{e^{ - yt}}\cos \left( {xt} \right)dt\,dx}  = } \int\limits_0^\infty  {\int {{e^{ - {t^2}/4}}{e^{ - yt}}\cos \left( {xt} \right)\,dx} \,dt},
$$
we can derive the following approximation
\begin{equation}\label{eq_8}
\begin{aligned}
{I_{{\rm{ext}}{\rm{.}}}}\left( {x,y} \right) &\approx \int\limits_0^\infty  {\int {\cos \left( {\frac{t}{{\sqrt 2 }}} \right){e^{ - yt}}\cos \left( {xt} \right)\,dx} \,dt} \\
 &  = \frac{1}{{2\sqrt \pi  }}\left( {\arctan \left( {\frac{{x + 1/\sqrt 2 }}{y}} \right) + \arctan \left( {\frac{{x - 1/\sqrt 2 }}{y}} \right)} \right).
\end{aligned}
\end{equation}

\subsection{Main internal domain}

Applying a new methodology of sampling based on incomplete cosine expansion of the sinc function we obtained the following approximation of the complex error function \cite{Abrarov2015, Abrarov2018a} (see also \cite{Mangaldan2021})
\begin{equation}\label{eq_9}
w\left( z \right) = {e^{ - {z^2}}}\left({1 - {\rm{erf}}\left( { - iz} \right)} \right) \approx \sum\limits_{m = 1}^M {\frac{{{a_m} + {b_m}\left( {z + i\varsigma /2} \right)}}{{c_m^2 - {{\left( {z + i\varsigma /2} \right)}^2}}}} ,
\end{equation}
where $z = x + iy$ is the complex argument, $N = 23$, $M = 23$, $h = 0.25$, $\varsigma  = 2.75$ are the adjustable parameters and
$$
{a_m} = \frac{{\sqrt \pi  \left( {m - 1/2} \right)}}{{2{M^2}h}}\sum\limits_{n =  - N}^N {{e^{{\varsigma ^2}/4 - {n^2}{h^2}}}\sin \left( {\frac{{\pi \left( {m - 1/2} \right)\left( {nh + \varsigma /2} \right)}}{{Mh}}} \right)},
$$
\[
{b_m} =  - \frac{i}{{M\sqrt \pi  }}\sum\limits_{n =  - N}^N {{e^{{\varsigma ^2}/4 - {n^2}{h^2}}}\cos \left( {\frac{{\pi \left( {m - 1/2} \right)\left( {nh + \varsigma /2} \right)}}{{Mh}}} \right)},
\]
$$
{c_m} = \frac{{\pi \left( {m - 1/2} \right)}}{{2Mh}},
$$
are the expansion coefficients. Due to rapid performance and high accuracy, the approximation \eqref{eq_9} may be used as an alternative to the Voigt/complex error function published in our earlier work (see equation (14) in \cite{Abrarov2011}) that is currently implemented in the latest version of the LBL atmospheric model MODTRAN \cite{Berk2017}.

The equation \eqref{eq_9} can be conveniently rearranged by defining the $\Omega $-function as
$$
\Omega \left( z \right) = \sum\limits_{m = 1}^M {\frac{{{a_m} + {b_m}z}}{{c_m^2 - {z^2}}}} 
$$
such that
$$
w\left( z \right) \approx \Omega \left( {z + i\varsigma /2} \right).
$$
Integrating now the $\Omega $-function with respect to variable $x$
$$
\bar I\left( {x,y} \right) = \int {\Omega \left( {x,y} \right)dx}
$$
leads to
\begin{equation}\label{eq_10}
\begin{aligned}
\bar I\left( {x,y} \right) =&\sum\limits_{m = 1}^M {\left[ {{\alpha _m}\left( {\arctan \left( {\frac{{{\gamma _m} - x}}{y}} \right) - \arctan \left( {\frac{{{\gamma _m} + x}}{y}} \right)} \right)} \right.} \\
&\left. { + {\beta _m}\left( {\ln \left( {\frac{{{{\left( {{\gamma _m} + x} \right)}^2} + {y^2}}}{{{{\left( {{\gamma _m} - x} \right)}^2} + {y^2}}}} \right)} \right)} \right],
\end{aligned}
\end{equation}
where the corresponding expansion coefficients are
$$
{\alpha _m} =  - \frac{1}{{2M\sqrt \pi  }}\sum\limits_{n =  - N}^N {{e^{{\varsigma ^2}/4 - {n^2}{h^2}}}\cos \left( {\frac{{\pi \left( {m - 1/2} \right)\left( {nh + \varsigma /2} \right)}}{{Mh}}} \right)},
$$
\[{\beta _m} =  - \frac{1}{{4M\sqrt \pi  }}\sum\limits_{n =  - N}^N {{e^{{\varsigma ^2}/4 - {n^2}{h^2}}}\sin \left( {\frac{{\pi \left( {m - 1/2} \right)\left( {nh + \varsigma /2} \right)}}{{Mh}}} \right)}
\]
and
$$
{\gamma _m} = \frac{{\pi \left( {m - 1/2} \right)}}{{2Mh}}.
$$
Changing the variable $z \to z + i\varsigma /2$ yields
\begin{equation}\label{eq_11}
{I_{{\rm{main}}}}\left( {x,y} \right) \approx \bar I\left( {x,y + \frac{\varsigma }{2}} \right).
\end{equation}

\subsection{Supplementary internal domain}

At $\left| x \right| \to 0$ the accuracy of the approximation \eqref{eq_11} deteriorates. In order to resolve this problem we can expand the cosine function into the Maclaurin series expansion
$$
\cos \left( {xt} \right) = 1 - \frac{{{x^2}{t^2}}}{{2!}} + \frac{{{x^4}{t^4}}}{{4!}} - \frac{{{x^6}{t^6}}}{{6!}} \ldots  \Rightarrow 1 + O\left( {{x^2}{t^2}} \right),
$$
where $O$ is the asymptotic notation (the Bachmann–Landau notation). This leads to
$$
K\left( {x,y} \right) \approx \frac{1}{{\sqrt 2 }}\int\limits_0^\infty  {{e^{ - {t^2}/4}}{e^{ - yt}}\left( {1 + O\left( {{x^2}{t^2}} \right)} \right)dt, \quad y \ge 0}, \quad \left| x \right| <  < 1
$$
or
$$
K\left( {x,y} \right) \approx \sqrt {\frac{\pi }{2}} {e^{{y^2}}}{\rm{erfc}}\left( y \right) + \varepsilon \left( {x,y} \right), \quad y \ge 0,\quad \left| x \right| <  < 1,
$$ such that
\[
\mathop {\lim }\limits_{x \to 0} \varepsilon \left( {x,y} \right) = 0
\]
and
$$
K\left( {0,y} \right) = \sqrt {\frac{\pi }{2}} {e^{{y^2}}}{\rm{erfc}}\left( y \right).
$$
Consequently, at small $x$ the error term $\varepsilon \left( {x,y} \right)$ is very close to zero. Therefore, we can write
$$
\int {\sqrt {\frac{\pi }{2}} {e^{{y^2}}}{\rm{erfc}}\left( y \right) + \varepsilon \left( {x,y} \right)dx \approx } \,x\left( {\sqrt {\frac{\pi }{2}} {e^{{y^2}}}{\rm{erfc}}\left( y \right) + \varepsilon \left( {x,y} \right)} \right), \qquad \left| x \right| <  < 1,
$$ 
or
$$
{I_{{\rm{sup}}{\rm{.}}}}\left( {x,y} \right) \approx x\,K\left( {0,y} \right), \qquad \left| x \right| <  < 1
$$
or
\begin{equation}\label{eq_12}
{I_{\rm{sup.}}}\left( {x,y} \right) \approx x\,{I_{{\rm{main}}}}\left( {\delta,y} \right)/\delta, \qquad \left| x \right| << 1,
\end{equation}
since from equation \eqref{eq_3} it follows that
$$
K\left( {0,y} \right) = \mathop {\lim }\limits_{\delta \to 0} \frac{{I\left( {0+\delta,y} \right)-I\left( {0,y} \right)}}{{\delta}}= \mathop {\lim }\limits_{\delta \to 0} \frac{{I\left( {0 + \delta,y} \right)}}{{\delta}}= \mathop {\lim }\limits_{\delta \to 0}K\left( {\delta,y} \right).
$$
In approximation \eqref{eq_12} we can take a small constant $\delta = 10^{-6}$.

\subsection{Arctangent function}

Equations \eqref{eq_8}, \eqref{eq_11} and \eqref{eq_12} employ the arctangent function. The built-in arctangent function in MATLAB is rapid enough in computation. However, as an alternative the following series expansion of the arctangent function that we derived in \cite{Abrarov2017} (see also \cite{Abrarov2018b})
$$
\arctan \left( \theta  \right) = 2\sum\limits_{n = 1}^\infty  {\frac{1}{{2n - 1}}\frac{{{a_n}\left( \theta  \right)}}{{a_n^2\left( \theta  \right) + b_n^2\left( \theta  \right)}}},
$$
where
\[
\begin{aligned}
&{a_1} = 2/\theta ,{b_1} = 1,\\
&{a_n}\left( \theta  \right) = \left( {1 - 4/{\theta ^2}} \right){a_{n - 1}}\left( \theta  \right) + 4{b_{n - 1}}/\theta,\\
&{b_n}\left( \theta  \right) = \left( {1 - 4/{\theta ^2}} \right){b_{n - 1}}\left( \theta  \right) - 4{a_{n - 1}}\left( \theta  \right)/\theta,
\end{aligned}
\]
can also be used.

\section{Results and discussion}

\subsection{Error analysis}

The SIVF code is implemented in the MATLAB according to the following scheme
$$
I\left( {x,y} \right) \approx \left\{ \begin{aligned}
&\left| {x + iy} \right| \ge 15, \hspace{2.75cm}  {I_{{\rm{ext.}}}}\left( {x,y} \right)\\
&\left| {x + iy} \right| < 15 \cap \left| x \right| \ge {10^{ - 6}}, \quad {I_{{\rm{main}}}}\left( {x,y} \right) \\
&\left| {x + iy} \right| < 15 \cap \left| x \right| < {10^{ - 6}}, \quad {I_{{\rm{sup.}}}}\left( {x,y} \right)
\end{aligned} \right.
$$
where ${I_{{\rm{ext}}{\rm{.}}}}\left( {x,y} \right)$, ${I_{{\rm{main}}}}\left( {x,y} \right)$ and ${I_{{\rm{sup.}}}}\left( {x,y} \right)$ are given by equations \eqref{eq_8}, \eqref{eq_10} and \eqref{eq_11}, respectively.

The accuracy of computation can be analyzed by using the relative error defined as
$$
\Delta  = \frac{{\left| {{I_{{\rm{ref}}{\rm{.}}}}\left( {x,y} \right) - I\left( {x,y} \right)} \right|}}{{{I_{{\rm{ref.}}}}\left( {x,y} \right)}},
$$
where $I_{{\rm{ref.}}}$ is the reference. Our algorithm is built to keep higher accuracy in the internal domain since for the external domain the high accuracy is not required.

\begin{figure}[ht]
\begin{center}
\includegraphics[width=20pc]{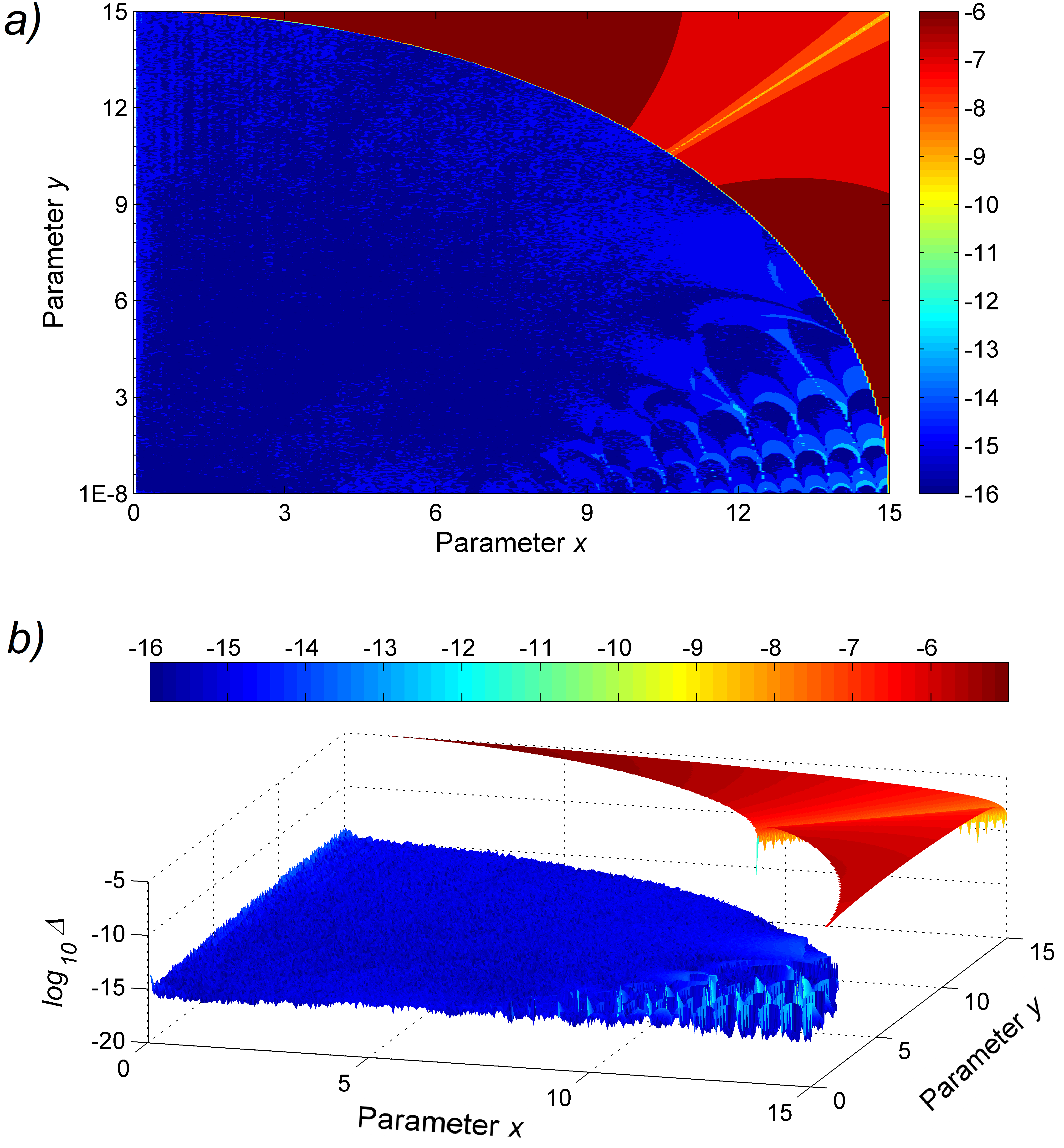}\hspace{2pc}%
\begin{minipage}[b]{22pc}
\vspace{0.3cm}
{\sffamily {\bf{Fig. 4.}} Relative error for the SIVF algorithm near main domain in log scale: a) contour plot and b) 3D plot.}
\end{minipage}
\end{center}
\end{figure}

Figures 4a and 4b show the relative error over the domain $0 \le x \le 15$ and ${10^{ - 8}} \le y \le 15$ in contour and 3D plots, respectively. As we can see, the main and supplementary subdomains within internal domain are highly accurate with relative error better than ${10^{ - 13}}$ (blue color area). Accuracy of the external domain is ${10^{ - 6}}$ (red color area).

Figure 5a and 5b illustrate the relative error over the domain $0 \le x \le 150$ and ${10^{ - 8}} \le y \le 150$ in contour and 3D plots, respectively. As we can see, the worst accuracy in the external domain is about ${10^{ - 6}}$. Thus, we can see that our algorithm provides an accuracy required for the HITRAN based applications.

It should be noted that a new methodology of computation based on the vectorized interpolation, proposed recently in our recent publication \cite{Abrarov2019}, can also be used for the SIVF algorithm to avoid unnecessary interpolation of the absorption coefficients that are typically implemented in conventional LBL algorithms \cite{Fomin1995, Sparks1997}.

\begin{figure}[ht]
\begin{center}
\includegraphics[width=20pc]{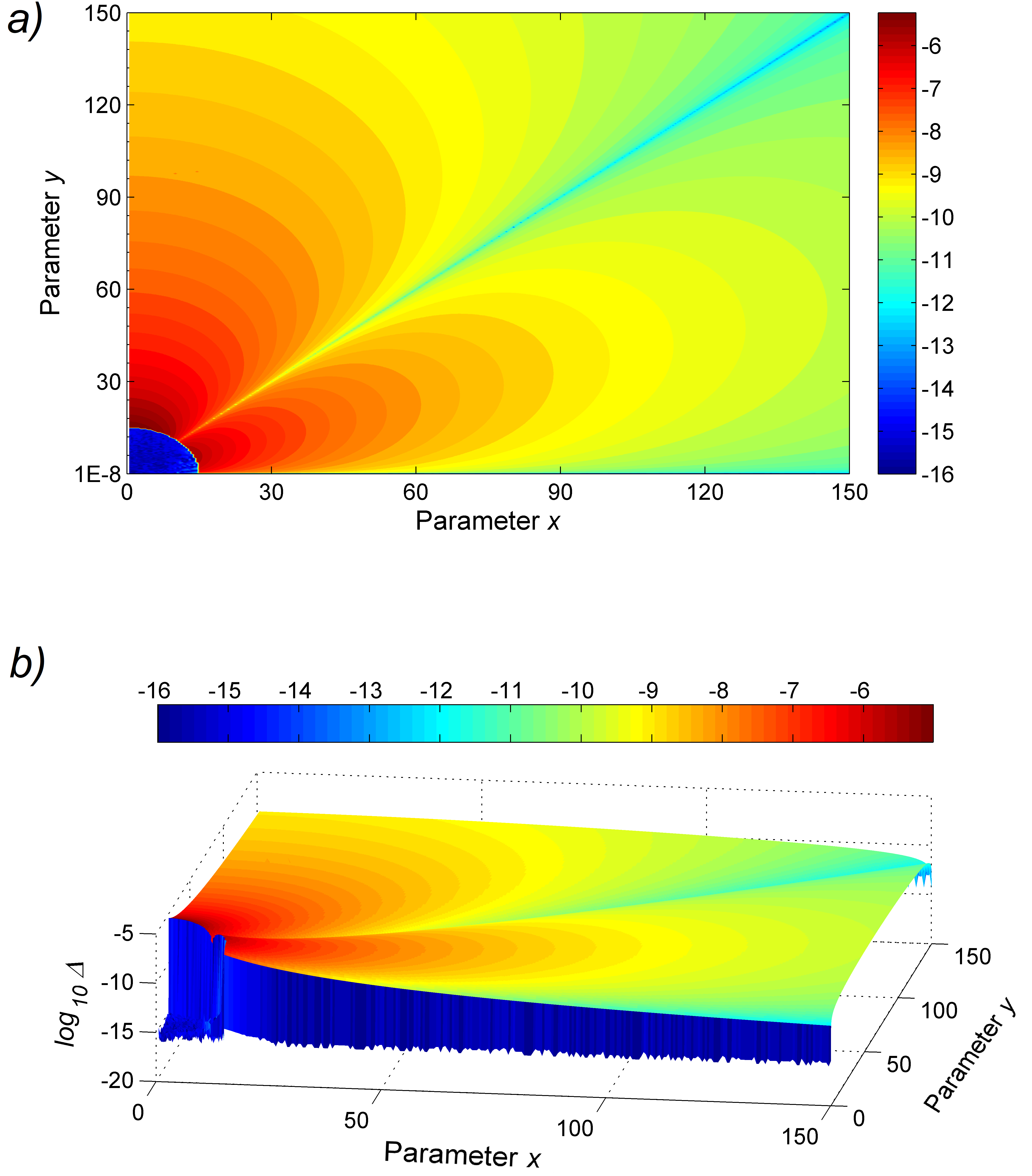}\hspace{2pc}%
\begin{minipage}[b]{22pc}
\vspace{0.3cm}
{\sffamily {\bf{Fig. 5.}} Relative error for the SIVF algorithm over main and external domains in log scale: a) contour plot and b) 3D plot.}
\end{minipage}
\end{center}
\end{figure}

\subsection{Application at low resolution}

Consider Fig. 6 showing the spectral radiance of $\rm{CO_2}$, $\rm{H_2O}$, $\rm{O_2}$ and $\rm{CH_4}$ greenhouse gases for $15 - 30\,km$ atmospheric column with 4 cells computed by regular Voigt function method at high spectral resolution $\Delta \nu = 0.0001\,c{m^{ - 1}}$ in the spectral range from $900\,c{m^{ - 1}}$ to $1000\,c{m^{ - 1}}$. All lines are present in this figure due to a high spectral resolution. However, at low spectral resolution $\Delta \nu = 0.1\,c{m^{ - 1}}$ many lines are missed while the shape of the spectral band is distorted. This can be seen from the Fig. 7. Application of SIVF instead of the Voigt function appeared to be more efficient at low spectral resolution. Specifically, as we can see from the Fig. 8 at $\Delta \nu = 0.1\,c{m^{ - 1}}$ none of the lines are missed while the shape of the spectral band is well-preserved.

%\newpage
\begin{figure}[ht]
\begin{center}
\includegraphics[width=22pc]{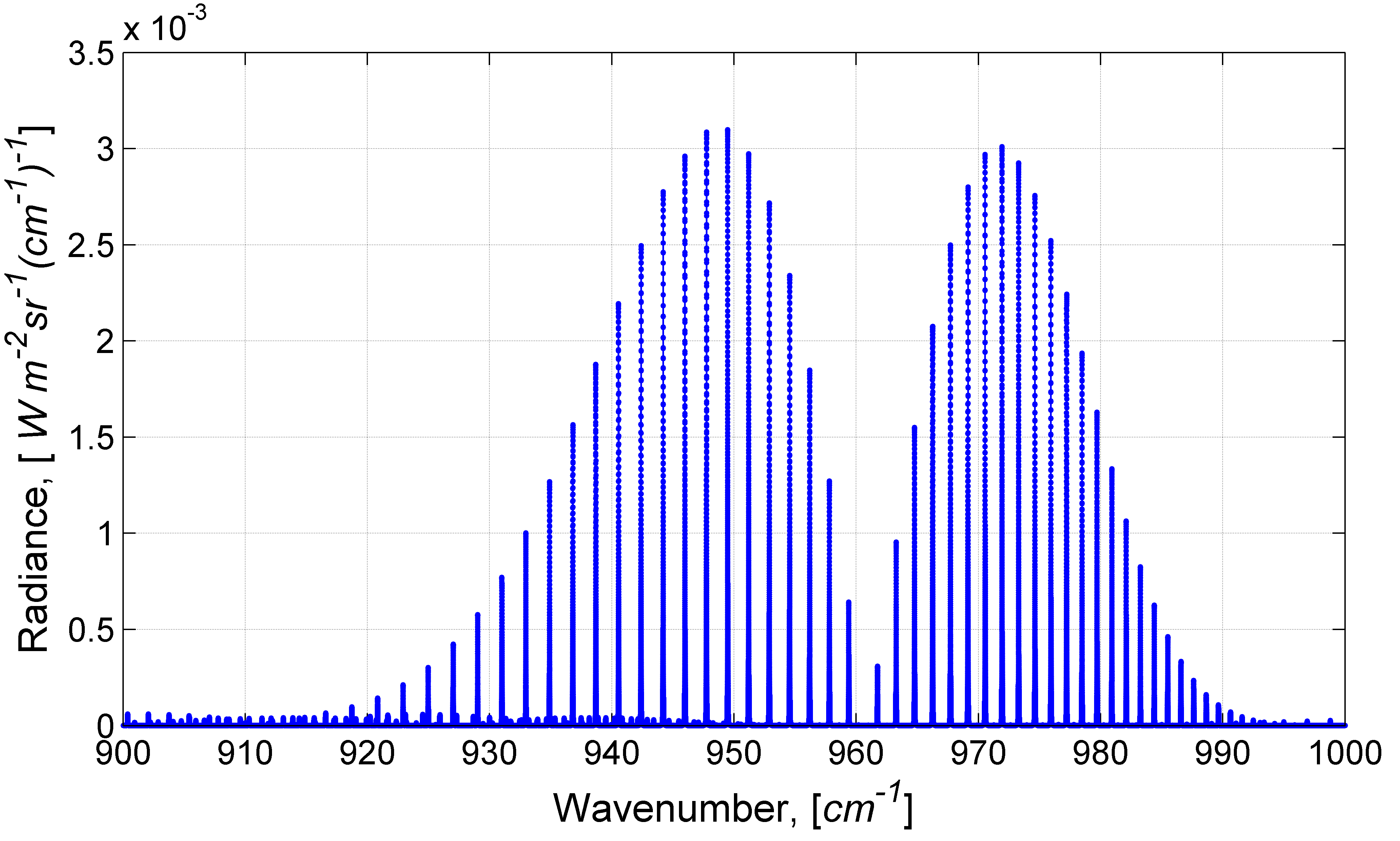}\hspace{2pc}%
\begin{minipage}[b]{24pc}
\vspace{0.3cm}
{\sffamily {\bf{Fig. 6.}} Spectral radiance of $\rm{CO_2}$, $\rm{H_2O}$, $\rm{O_2}$ and $\rm{CH_4}$ greenhouse gases for $15 - 30\,km$ atmospheric layer ($4$ cells) calculated at high spectral resolution (step $\Delta \nu = 0.0001\,c{m^{ - 1}}$) by conventional method in the spectral range from $900\,c{m^{ - 1}}$ to $1000\,c{m^{ - 1}}$.}
\end{minipage}
\end{center}
\end{figure}

\begin{figure}
\begin{center}
\includegraphics[width=22pc]{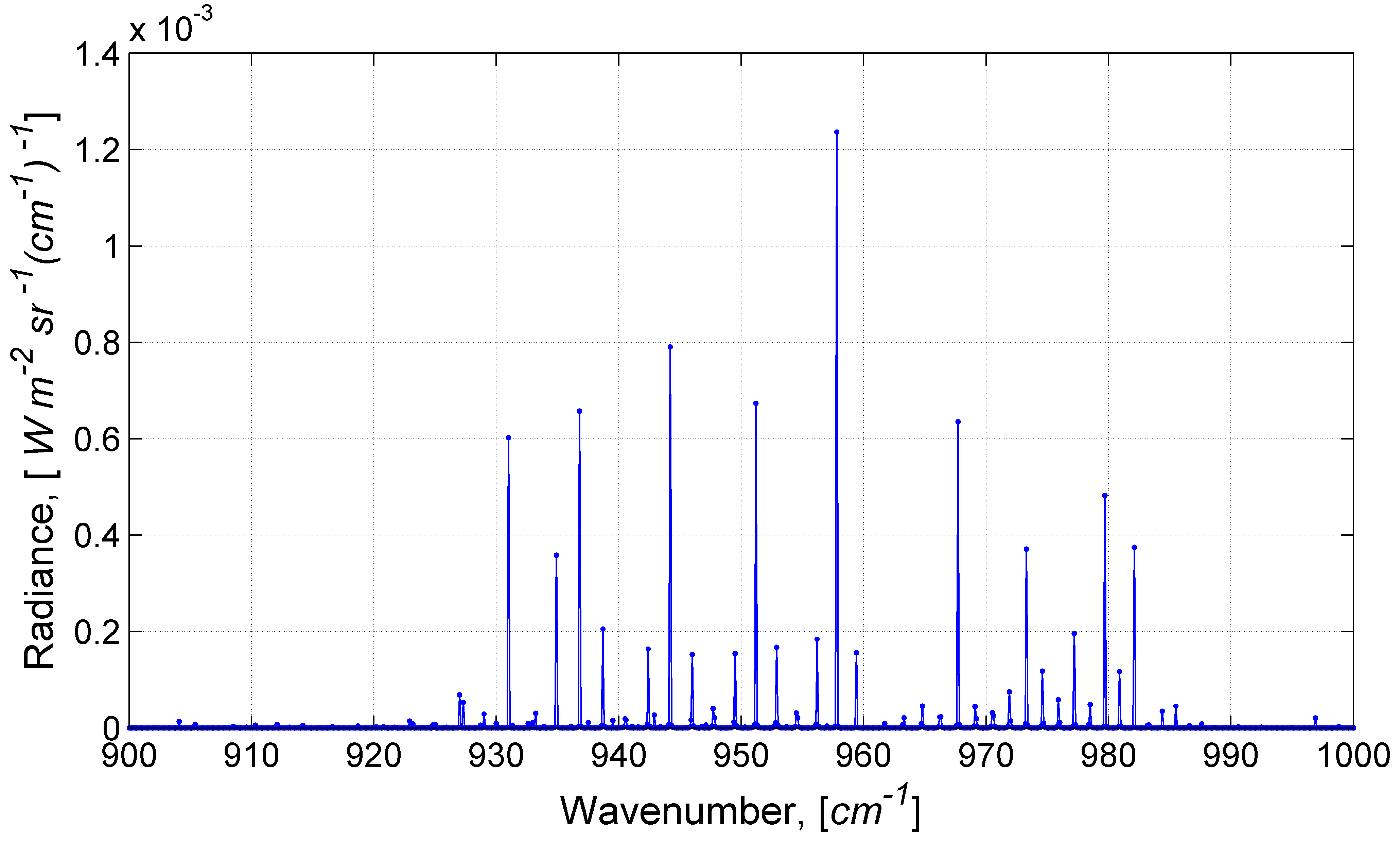}\hspace{2pc}%
\begin{minipage}[b]{24pc}
\vspace{0.3cm}
{\sffamily {\bf{Fig. 7.}} Conventional method of computation with step $\Delta \nu = 0.1\,c{m^{ - 1}}$. Most lines are missed and the shape is completely distorted (loss of information).}
\end{minipage}
\end{center}
\end{figure}

Figure 9 shows the total band radiances of $\rm{CO_2}$, $\rm{H_2O}$, $\rm{O_2}$ and $\rm{CH_4}$ greenhouse gases in the band from $900\,c{m^{ - 1}}$ to $1000\,c{m^{ - 1}}$ computed by the regular Voigt function at high (blue curve) and low (red curve) spectral resolutions at $\Delta \nu = 0.0001\,c{m^{ - 1}}$ and $\Delta \nu = 0.1\,c{m^{ - 1}}$, respectively. As we can see from this figure, the total band radiance shown by red curve is not accurate. In particular, a strong discrepancy is observed near $960\,c{m^{ - 1}}$. In contrast, the total band radiance computed by the SIVF is accurate. This can be seen from the Fig. 10 showing the total band radiances of $\rm{CO_2}$, $\rm{H_2O}$, $\rm{O_2}$ and $\rm{CH_4}$ greenhouse gases in the band from $900\,c{m^{ - 1}}$ to $1000\,c{m^{ - 1}}$ computed by the regular Voigt function at high (blue curve) and SIVF at low (red curve) spectral resolutions at $\Delta \nu = 0.0001\,c{m^{ - 1}}$ and $\Delta \nu = 0.1\,c{m^{ - 1}}$, respectively. The stepwise behavior of total band radiance is observed due to cumulative contribution of individual lines from the $\rm{CO_2}$, $\rm{H_2O}$, $\rm{O_2}$ and $\rm{CH_4}$ greenhouse gases.

\begin{figure}
\begin{center}
\includegraphics[width=22pc]{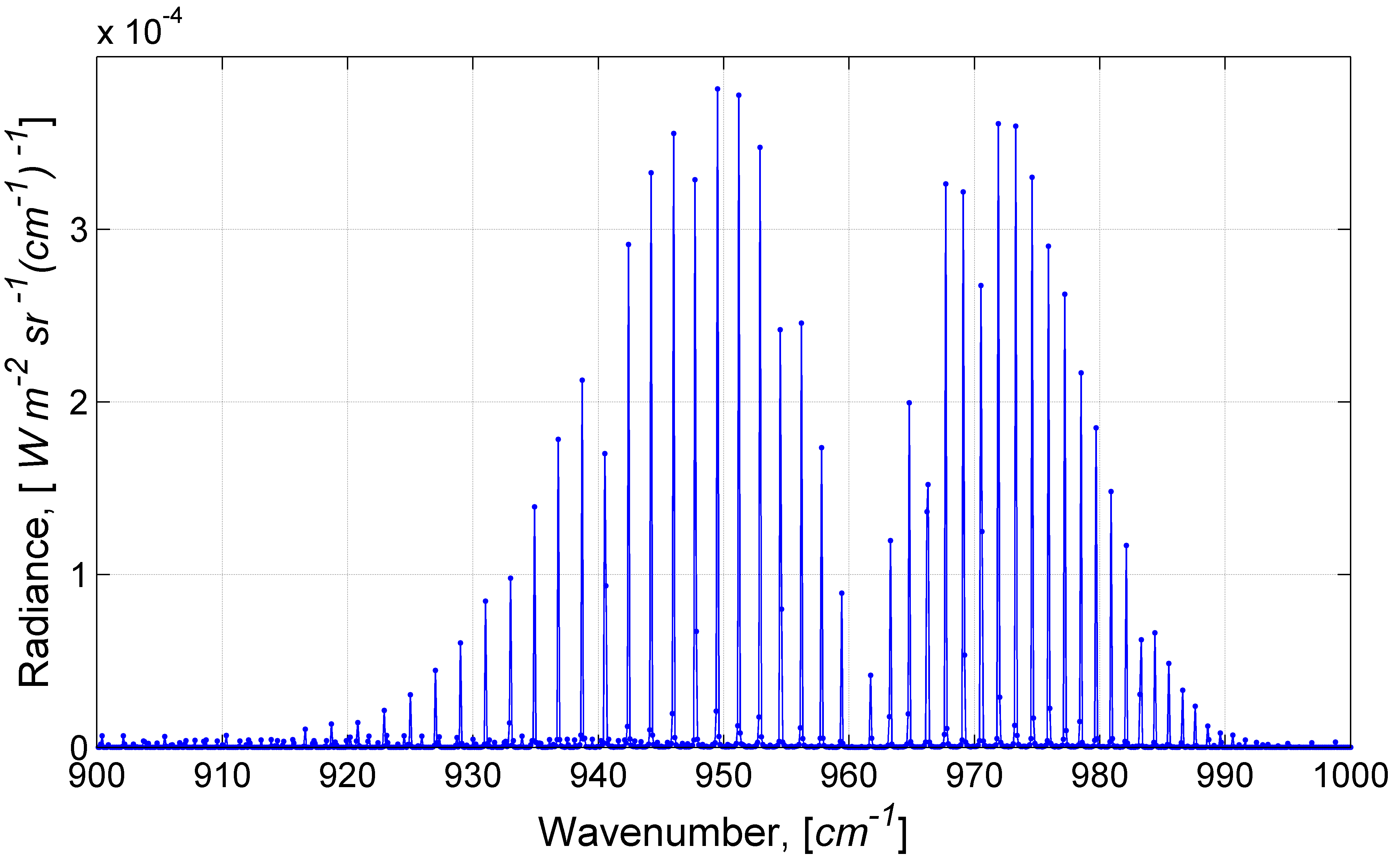}\hspace{2pc}%
\begin{minipage}[b]{24pc}
\vspace{0.3cm}
{\sffamily {\bf{Fig. 8.}} The SIVF method of computation with step $\Delta \nu = 0.1\,c{m^{ - 1}}$. Despite low resolution, the shape of the spectral radiance is well-preserved and all lines are present.}
\end{minipage}
\end{center}
\end{figure}

Figure 11 illustrates the absolute difference for the total band radiance when the conventional Voigt function (blue curve) and SIVF (red curve) are used in computation at low spectral resolution $\Delta \nu = 0.1\,c{m^{ - 1}}$. As we can see from this figure, the conventional Voigt function has a largest discrepancy almost reaching $1.4 \times {10^{ - 14}}\,W{m^{ - 2}}s{r^{ - 1}}$ at around $960\,c{m^{ - 1}}$. However, in contrast to the conventional method, the absolute difference in case of SIVF is considerably smaller by order of the magnitude $3 \times {10^{ - 15}}\,W{m^{ - 2}}s{r^{ - 1}}$  and observed at about $975\,c{m^{ - 1}}$. Thus, we can see that despite low resolution $\Delta \nu = 0.1\,c{m^{ - 1}}$, the SIVF method can be significantly more accurate in computation of the total band radiance.

\begin{figure}
\begin{center}
\includegraphics[width=22pc]{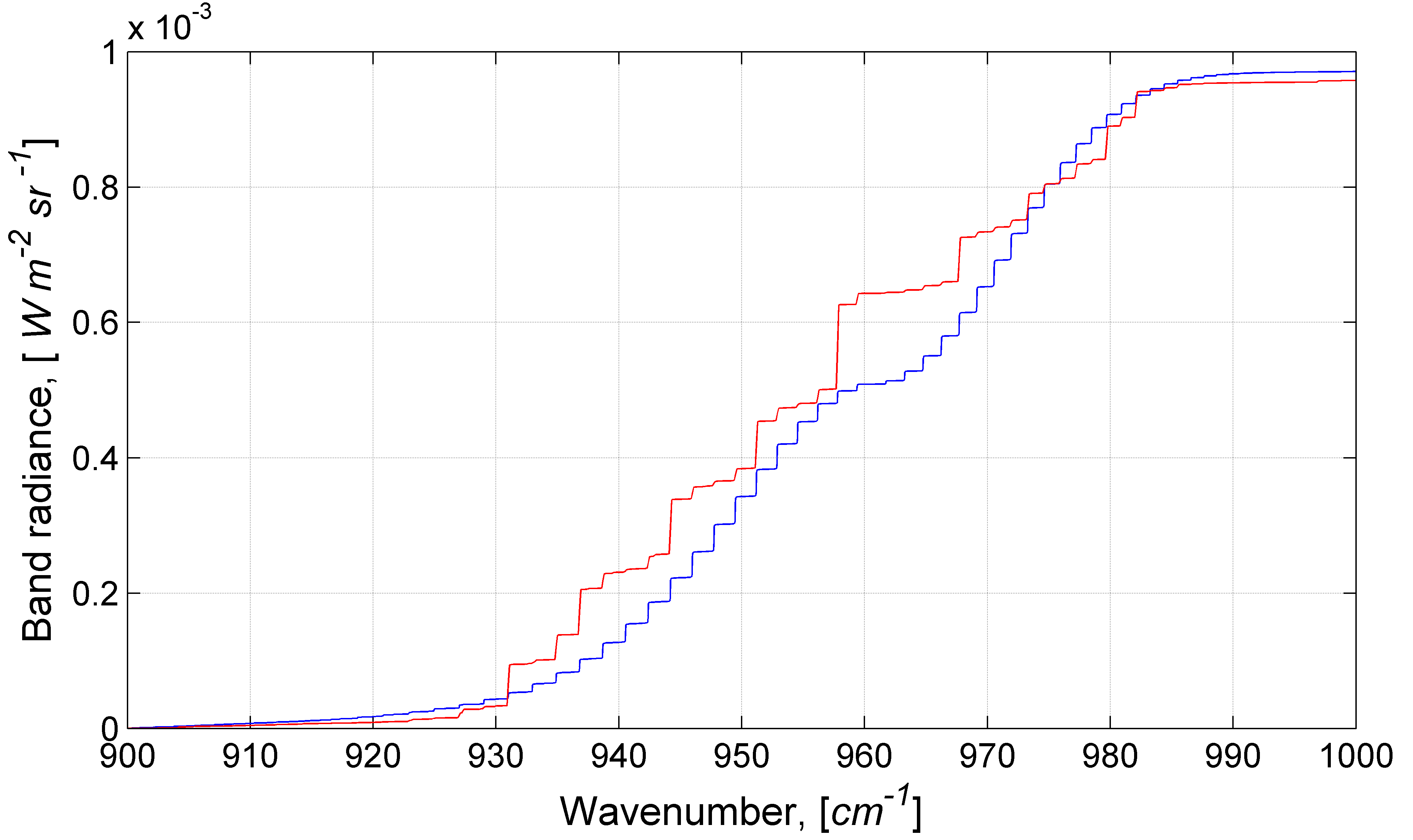}\hspace{2pc}%
\begin{minipage}[b]{24pc}
\vspace{0.3cm}
{\sffamily {\bf{Fig. 9.}} Total band radiance at fixed ${\nu_{\min }} = 900\,c{m^{ - 1}}$ and increasing ${\nu_{\max }}$ from $900\,c{m^{ - 1}}$ to $1000\,c{m^{ - 1}}$. The largest discrepancy is observed at about $960\,c{m^{ - 1}}$.
}
\end{minipage}
\end{center}
\end{figure}

\begin{figure}
\begin{center}
\includegraphics[width=22pc]{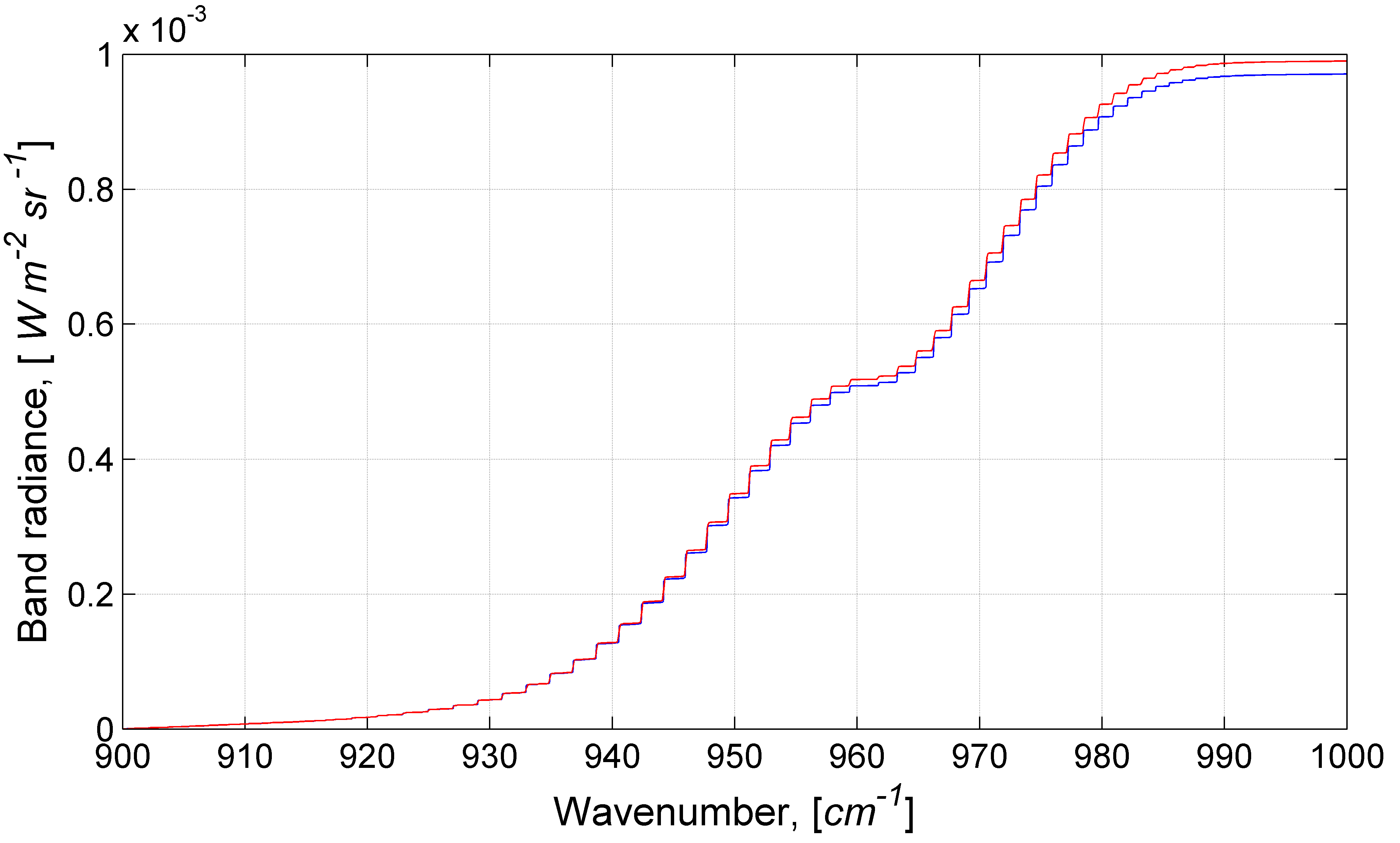}\hspace{2pc}%
\begin{minipage}[b]{24pc}
\vspace{0.3cm}
{\sffamily {\bf{Fig. 10.}} Total band radiance at fixed ${\nu_{\min }} = 900\,c{m^{ - 1}}$ and increasing ${\nu_{\max }}$ from $900\,c{m^{ - 1}}$ to $1000\,c{m^{ - 1}}$. Despite low resolution, there is a good match between two curves.}
\end{minipage}
\end{center}
\end{figure}

\begin{figure}
\begin{center}
\includegraphics[width=22pc]{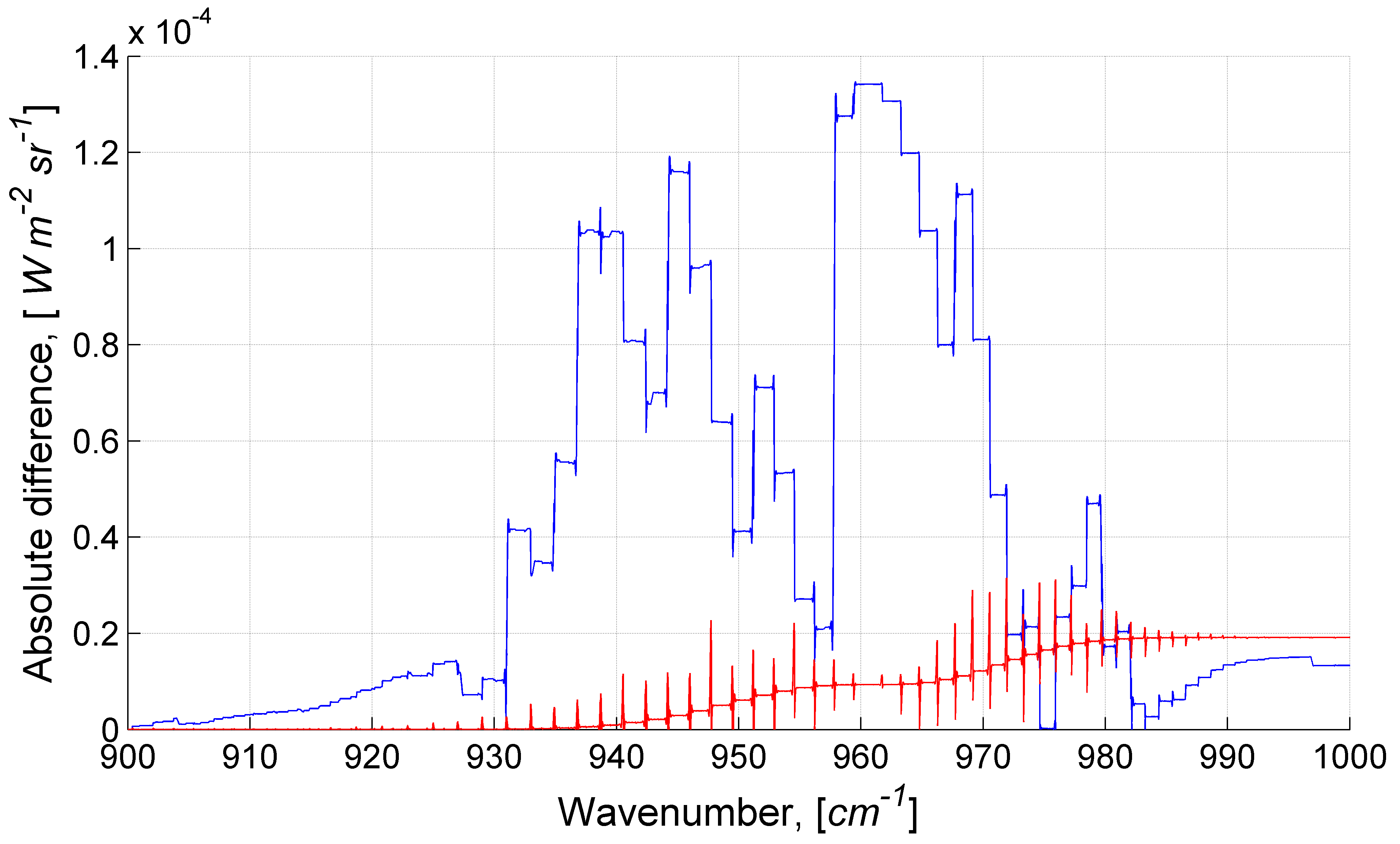}\hspace{2pc}%
\begin{minipage}[b]{24pc}
\vspace{0.3cm}
{\sffamily {\bf{Fig. 11.}} Absolute difference computed at fixed ${\nu_{\min }} = 900\,c{m^{ - 1}}$ and increasing ${\nu_{\max }}$ from $900\,c{m^{ - 1}}$ to $1000\,c{m^{ - 1}}$.  Discrepancy in SIVF method by about one order of the magnitude is smaller.}
\end{minipage}
\end{center}
\end{figure}

The application of the SIVF may also be used in detection of clouds and seasonal forest fires by using a new Radiance Enhancement (RE) method that was reported in our recent publications \cite{Siddiqui2017, Siddiqui2020}. This RE method provides probability of finding cloud scene by using infra-red data from the space orbiting Argus 1000 micro-spectrometer. The lower spectral resolution can also be used to accelerate the data retrieval, analysis and interpretation for the detection of the cloud scenes as well as the seasonal forest fires \cite{Siddiqui2020}.

\section{Conclusion}

We show that application of the SIVF can be used to retain its accuracy in computation of the total band radiance at low spectral resolution. This is possible to achieve since SIVF mathematically represents the mean value integral of the Voigt function that it accounts for the area under the curve of the Voigt function. Due to reduced spectral resolution application of the SIVF may be promising for accelerated line-by-line computation in atmospheric models based on the HITRAN molecular spectroscopic database \cite{Hill2016}. This approach may be particularly efficient to implement a rapid retrieval algorithm for the greenhouse gases from the NIR space data collected by space-orbiting micro-spectrometers like Argus 1000 for their operation in a real-time mode. The real-time mode operation of the micro-spectrometers is required for instant decision making during flight for more efficient data collection from space.

\section*{Acknowledgments}

This work is supported by the National Science and Engineering Research Council of Canada, Thoth Technology Inc., Epic Climate Green (ECG) Inc., York University and Epic College of Technology.

\bigskip
%\newpage

\end{document}